\documentstyle[12pt]{article}
\begin{document}
\begin{titlepage}
\title{\Large\bf GEODESIC LINES IN THE GRAVITATIONAL FIELD 
OF NONLINEAR COSMIC STRINGS}
\author{{\bf L. M. Chechin$^1$  and T. B. Omarov}\\
{\normalsize Astrophysics Institute, National Academy of Sciences,}\\
{\normalsize Almaty, 480068, Kazakstan}}
\date{\small Received 24 March 2000}
\maketitle
\abstract{We briefly review the equations of motion and the space-time
interval due to the nonlinear cosmic string that have been derived in 
ref. [3] for the first time. 
The different types of isotropic and nonisotropic
geodesic lines in the gravitational field of nonlinear cosmic string 
have been analyzed in detail}.

\vfill
$^1$E-mail: chel\mbox{@}afi.academ.alma-ata.su\\
.
\end{titlepage}

\section{Introduction}

It is well known that cosmic strings were appeared at the very early stages
of the Universe evolution. The cosmic string "sources" are the scalar fields
that were, in turn, produced due to the vacuum phase transitions [1].
One of the lagrangians that describes these fields is the Higgs's lagrangian for the
complex scalar field $\chi(x)$,
\begin{equation}
{\cal L} = \partial_{\alpha}\chi^{\ast}\partial^{\alpha}\chi + m^2\chi^{\ast}
\chi - \lambda(\chi^{\ast}\chi)^2,
\end{equation}
where $\lambda$ is coupling constant describing the field self-interaction.
The lagrangian is essentially nonlinear. This is the reason that,
in general, cosmic strings are nonlinear as well.

To describe the strings, let us study the thread-like matter that consists 
of the set of infinitely thin crossless threads and a fully continuous 
spacetime. The energy-momentum tensor of this thread-like matter 
can be written as
\begin{equation}
T^{\alpha\beta} = \mu u^{\alpha}u^{\beta} - t^{\alpha\beta}.
\end{equation}
The second term can be interpreted as the stress tensor that appeared due to
the oscillations of each thread. These oscillations occur in view of the
string's tension, or due to elastic force. We should stress, in a complete 
analogy with the classical mechanics [2], that the elastic force must include
not only linear term but also a nonlinear one. By virtue of the
nonlinear term, the stress tensor of the thread-like matter has 
the following form:
\begin{equation}
t^{\alpha\beta} = \mu l^{\alpha}l^{\beta}(1 - \frac{\varepsilon^2}{2}l^{\gamma}
l_{\gamma}),
\end{equation}
where $\displaystyle\frac{\varepsilon^2}{2}$ is the corresponding coupling 
constant of the order of $\lambda$. Inserting eqs. (2) and (3) into 
the energy-momentum consrevational law and using the standard notation, 
we get the equations of motion of nonlinear cosmic string [3],
\begin{equation}
\ddot{x^{\alpha}} - x''^{\alpha}(1 - \frac{\varepsilon^2}{2}x'^{\gamma}x'_{\gamma})
= 0.
\end{equation}
The nonlinear character of cosmic string leads to the appearance of some
additional terms in the metric coefficients that describe the spacetime 
interval in the vicinity of such a string.

Indeed, inserting eqs. (2) and (3) into the Einstein's gravitational field
equations, putting that cosmic string along the $z$-axis and specifying
the results of [3], we get the 4-dimensional interval in quasi-cartesian 
coordinates in the following form:
$$ds^2 = \big(1 - 2\gamma\mu\varepsilon^2\ln\frac{r}{r_0}\big){dx^0}^2 - $$
\begin{equation}
\big[1 - 8\gamma\mu(1 + \frac{\varepsilon^2}{4})\ln\frac{r}{r_0}\big]
(d{x^1}^2 + d{x^2}^2) -
\end{equation}
$$[1 + 2\gamma\mu\varepsilon^2\ln\frac{r}{r_0}]d{x^3}^2.$$
For the case $\varepsilon = 0$, the expression (5) recovers  
the standard Vilenkin's spacetime interval [4].

Deducing of Vilenkin's metric ansatz for the rectilinear massive cosmic
string has stimulated a number of research papers devoted their cosmological 
and astrophysical applications; e.g., motion of a test particle, 
nonrelativistic [4] and relativistic [5], and light ray propagation 
in the solitary cosmic string background.

Moreover, in all subsequent generalizations of this metric the behaviour of
geodesic lines has been examined as well. Particularly, 
the dynamics of test particle in the gravitational field of cosmic 
string with kink has been studied in [6]; 
light deflection caused by a cosmic string carrying a wave pulse was 
investigated in [7]; 
the motion of light rays and test particles in the gravitational field 
of a semiclassical cosmic string was considered in [8]; 
the dynamics of test particles in the gravitational field of cosmic 
string passing through a black hole, through a domain wall was 
analyzed in [9]; 
the influence of the current-carrying cosmic string background on 
the light rays propagation was treated in [10].

In this paper, in accordance to the general logic of investigation 
of the new types of cosmic string metrics, briefly discussed above, 
we study geodesics lines in the gravitational field of nonlinear 
cosmic string.

\section{General form of a geodesic line in the nonlinear cosmic string
background}

Let us write down the spacetime interval (5) in quasi-cylindrical 
coordinates,
$$ds^2 = \big(1 - 2\gamma\varepsilon^2\ln\frac{r}{r_0}\big){dx^0}^2 -$$
\begin{equation}
\big(1 - 8\gamma\mu(1 + \frac{\varepsilon^2}{4})\ln\frac{r}{r_0}\big)
(dr^2 + r^2{d\varphi}^2) -
\end{equation}
$$\big(1 + 2\gamma\mu\varepsilon^2\ln\frac{r}{r_0}\big)dz^2.$$
Upon the admited coordinate transformations,
\begin{equation}
\left.
\begin{array}{ccc}
\bar{x}^0& = &x^0,\\
\bar{r}& = &r\big\{1 + 4\gamma\mu(1 + \displaystyle\frac{\varepsilon^2}{4})
(1 - \ln\displaystyle\frac{r}{r_0})\bigr\},\\
\bar{\varphi}& = &\varphi,\\
\bar{z}& = &z
\end{array}
\right\}
\end{equation}
and retaining corrections terms of the order of $\gamma\mu\varepsilon^2$, 
the expression (6) can be written in the form
$$ds^2 = \big(1 - 2\gamma\mu\varepsilon^2\ln\frac{\bar{r}}{\bar{r_0}}\big){d\bar{x^0}}^2 -
$$
\begin{equation}
{d\bar{r}}^2 - {\bar{r}}^2\big[1 - 8\gamma\mu\big](1 + \frac{\varepsilon^2}{4})
{d\bar{\varphi}}^2 - {d\bar{z}}^2.
\end{equation}
Introducing the new angle variable,
\begin{equation}
\bar{\Phi} = \big[1 - 4\gamma\mu\big](1 + \frac{\varepsilon^2}{4})\bar{\varphi}
\end{equation}
and omitting, for simplify, the minus sign at all the coordinates,
it is easy to verify that the interval (8) looks like 
a quasi-newtonian one. Namely, 
$$ds^2 = \big(1 - 2\gamma\mu\varepsilon^2\ln\frac{r}{r_0}\big){dx^0}^2 - $$
\begin{equation}
{dr}^2 - r^2{d\Phi}^2 - {dz}^2.
\end{equation}
We have accounted that the angle $\Phi$ varies in accord to
\begin{equation}
0 \leq \Phi \leq 2\pi b,
\end{equation}
where  $b = [1 - 4\gamma\mu](1 + \displaystyle\frac{\varepsilon^2}{4})$ .
The above form is the reason why the interval (10) describes 
the quasi-newtonian modified conical spacetime.

The direct procedure of finding lagrangian from the arbitrary spacetime 
interval has been presented in [11]. Using this procedure, 
for the test particle that moves in $z = const$ plane in the gravitational 
field with metric (10), we get
$$L = \big(1 - 2\gamma\mu\varepsilon^2\ln\frac{r}{r_0}\big)\big(\frac{dx^0}{ds}\bigr)^2 - $$
\begin{equation}
\big(\frac{dr}{ds}\big)^2 - r^2\big(\frac{d\Phi}{ds}\big)^2.
\end{equation}
As this lagrangian does not explicitly depend neither on $x^0$, 
nor on $\Phi$, then the derivatives of $L$ with respect to $dx^0/ds$ 
and $d\Phi/ds$ give two first integrals of motion: the total energy,
\begin{equation}
\big(1 - 2\gamma\mu\varepsilon^2\ln\frac{r}{r_0}\big)\frac{dx^0}{ds}
 = C_1  = E,
\end{equation}
and the total angular momentum,
\begin{equation}
r^2\frac{d\Phi}{ds} = C_2 = M,
\end{equation}
respectively.

Moreover, as the lagrangian (12) does not explicitly depend on $s$ too, 
then another integral of motion exists, namely,
$$\big(1 - 2\gamma\mu\varepsilon^2\ln\frac{r}{r_0}\big)\big(\frac{dx^0}{ds}\big)^2 - $$
\begin{equation}
\big(\frac{dr}{ds}\big)^2 - r^2\big(\frac{d\Phi}{ds}\big)^2 = C_3 = C.
\end{equation}
So, from eqs. (13) and (14) we have the 4-velocity components, 
\begin{equation}
\frac{dx^0}{ds} = \frac{E}{\displaystyle {1 - 2\gamma\mu\varepsilon^2\ln\frac{r}{r_0}}},
\end{equation}
\begin{equation}
\frac{d\Phi}{ds} = \frac{M}{r^2}.
\end{equation}

Substituting eqs. (16) and (17) into (15) and introducing 
new variable $u = 1/r$, we obtain the general form of the equations
of geodesics in the nonlinear cosmic string background, 
\begin{equation}
\bigl(\frac{du}{d\Phi}\bigr)^2 + u^2 = - \frac{C}{M^2} + \frac{E^2}{M^2\displaystyle{(1 + 
2\gamma\mu
\varepsilon^2\ln r_0u)}}.
\end{equation}

\section{Null geodesic lines in the nonlinear cosmic string background}

Let us consider the basic types of the null geodesic lines in the nonlinear cosmic
string spacetime characterized by the interval (10).

{\it a) Hyperbolic light rays propagation}. For any null geodesic line the
constant $C$ is equal to zero. Then from (18) we have, approximately, 
\begin{equation}
\bigl(\frac{du}{d\Phi}\bigr)^2 + u^2 = \frac{E^2}{M^2}(1 - 2\gamma\mu\varepsilon^2
\ln{r_0u}).
\end{equation}
Differentiating the above expression with respect to $\Phi$ we get
\begin{equation}
\frac{d^2u}{d{\Phi}^2} + u = - \gamma\mu\varepsilon^2\frac{E^2}{M^2}\frac{1}{u}.
\end{equation}
We look for the solution of this equation in the form
\begin{equation}
u = u_0 + u_1 + \cdots,
\end{equation}
where $u_0$ describes the free motion while $u_1$ describes 
a perturbation, up to the first order accuracy.

For the nonperturbed motion we have simple trajectory.
Namely, the equation 
\begin{equation}
\frac{d^2 u_0}{d{\Phi}^2} + u_0 = 0
\end{equation}
has the solution
\begin{equation}
u_0 = \frac{E}{M}\cos\Phi.
\end{equation}
It describes the straight-line trajectory that passes through 
the coordinate origin at the distance $\displaystyle\frac{M}{E}$.

Next, for the perturbed motion we can write the equation of trajectory as
\begin{equation}
\frac{d^2 u_1}{d{\Phi}^2} + u_1 = - \gamma\mu\varepsilon^2\frac{E}{M}\cos^{-1}\Phi,
\end{equation}
and its solution is
\begin{equation}
u_1 = - \gamma\mu\varepsilon^2 \frac{E}{M}(\Phi\sin\Phi + \cos\Phi \ln\cos\Phi).
\end{equation}
Hence, the general solution of the equation of motion (18) is
$$u = \frac{E}{M}(\cos\Phi - \gamma\mu\varepsilon^2\Phi\sin\Phi) - $$
\begin{equation}
\gamma\mu\varepsilon^2\frac{E}{M} \cos\Phi\ln\cos\Phi.
\end{equation}

Now, it is easy to find the light deflection angle from eq. (26). 
We note that in the case of straight-line trajectory
the polar angle varies from 
$-\displaystyle\frac{\pi}{2}b$ to $\displaystyle\frac{\pi}{2}b$. 
That is why the radius-vector can be rotated up to the full angle $\pi b$. 
The angle values $\Phi = \pm\displaystyle\frac{\pi}{2}b$ give us 
$u = 0$, i.e. $r = \pm\infty$.
All the above relations are valid for the nonperturbed trajectory (22). 
For the perturbed trajectory (24), the same angle values lead
to a small, but nonzero, value of $u$. 
Therefore, the last term is equal to zero if 
$\Phi > \pm\displaystyle\frac{\pi}{2}b$.

So, let us denote the angle at which $u = 0$ 
as $\pm\displaystyle\frac{\pi}{2}b + \delta$, 
where $\delta$ is a small angle. 
Substituting it into the eq. (26) we obtain, approximately,
\begin{equation}
\delta - \gamma\mu\varepsilon^2 \pi/2 = 0.
\end{equation}
Therefore, the angle $\delta$ is found as
\begin{equation}
\delta = \gamma\mu\varepsilon^2 \pi/2
\end{equation}
and the full deflection angle of the light rays in the nonlinear 
cosmic string background is
\begin{equation}
\Delta\Theta = 2\delta = \pi\gamma\mu\varepsilon^2.
\end{equation}

{\it b) Radial light rays propagation}. In this case from eq. (15) 
we approximately have
\begin{equation}
dx^0 = (1 + \gamma\mu\varepsilon^2\ln\frac{r}{r_0})dr.
\end{equation}
Integrating this expression from $0$ to arbitrary point $r$ 
gives us
\begin{equation}
x^0 = r\bigl[1 - \gamma\mu\varepsilon^2(1 - \ln\frac{r}{r_0})\bigr].
\end{equation}
We can use this relation to calculate the Shapiro's effect 
in the gravitational field of nonlinear cosmic string. 
Indeed, the lapse time corresponding to the nonlinearity 
of cosmic string can be easily obtained, 
\begin{equation}
\Delta x^0_{ncs} = \gamma\mu\varepsilon^2 (r_2 - r_1)\ln\frac{r_2 - r_1}{r_0}.
\end{equation}
To estimate the magnitude of the effect let us suppose that 
the nonlinear cosmic string is placed between Earth and Mercury,
in the plane which is normal to the planets plane of motion. In this
case, $r_2 - r_1 \approx 1,4  a.u.\approx 2,2 10^{11} m$. 
Assuming $\gamma\mu \sim 10^{-6}$ and $r_0 \sim 10^{-30}  m$, 
that refers to string line mass density and Compton
wavelength corresponding to GUT models, accordingly, 
$\varepsilon \sim 10^{- 7}$ [3], we get 
$\Delta t_{ncs} \sim 10^{- 15} sec$. 
One can see that this time interval is too small 
to be detected experimentally; 
note that Shapiro's effect due to gravitational field 
of the Sun is of the order of $10^{- 4} sec$.

\section{Nonisotropic geodesic lines in the gravitational field 
of nonlinear cosmic string}

Finally, let us consider the main types of timelike geodesic lines in 
the nonlinear cosmic string spacetime.

{\it a) Hyperbolic motion}. For the case of 
nonisotropic timelike geodesic lines, we have $C = 1$.
Inserting this value into eq. (18) we obtain, within the adopted 
accuracy,
$$\bigl(\frac{du}{d\phi}\bigr)^2 + u^2 = - \frac{1}{M^2} + $$
\begin{equation}
\frac{E^2}{M^2}(1 - 2\gamma\mu\varepsilon^2\ln r_0u).
\end{equation}
Formally, this equation differs from that of the isotropic geodesic line 
(19). However, differentiating it with respect to $\Phi$ leads to 
the equation that fully coincides with eq. (20),
\begin{equation}
\frac{d^2u}{d{\Phi}^2} + u = - \gamma\mu\varepsilon^2\frac{E^2}{M^2}\frac{1}{u}.
\end{equation}
In the following, we deal only with the nonperiodic terms in eq. (26). 
Therefore, the resulting solution is
$$u = \frac{E}{M}(\cos\Phi - \gamma\mu\varepsilon^2\Phi\sin\Phi) \approx $$
\begin{equation}
\frac{E}{M}\cos(1 -\gamma\mu\varepsilon^2)\Phi.
\end{equation}
It is easy to see that the above timelike trajectory 
is a little bit differerent from the straightlike one, eq. (22). 
This difference is described by the angle
\begin{equation}
\Delta\Theta = \pi\gamma\mu\varepsilon^2
\end{equation}
between the asymptotics of trajectories. 
This expression is completely equivalent to the angle (29) 
that had been calculated above. We note that in the case
of linear cosmic srting the deflection angles for test particle 
and light ray are equal to each other, too [9].

{\it b) Radial motion}. For this type of motion $\Phi = const$. 
Hence, from eq. (15) we get
$$v = \frac{dr}{dx^0} = \pm \big(\frac{E^2 - 1}{E^2}\bigr)^{1/2}\big(1 - $$
\begin{equation}
\frac{2 - E^2}{1 - E^2}\gamma\mu\varepsilon^2\ln\frac{r}{r_0}\bigr),
\end{equation}
which is the particle's free-fall velocity determined by clock 
of infinitely far observer.

The local motionless observer determs the velocity in a
different way,
$$w = \frac{dr}{d\tau} = \frac{dr}{dx^0}(1 + \gamma\mu\varepsilon^2\ln\frac{r}{r_0}) = $$
\begin{equation}
\bigl(\frac{E^2 - 1}{E^2}\bigr)^{1/2}\bigl(1 - \frac{1}{1 - E^2}\gamma\mu
\varepsilon^2\ln\frac{r}{r_0}\bigr).
\end{equation}
If $r \to r_0$, i.e. on the surface of nonlinear cosmic string, then  
both of the velocities, (37) and (38), are equal to their newtonian 
expression $v_n =(\displaystyle\frac{E^2 - 1}{E^2})^{1/2}$. 
The last relation means that the  velocity is constant,
due to the free motion in flat space.

It should be pointed out that the second term in eq. (37) may be 
equal to zero, at some conditions. 
Namely, for the specific value of the radial coordinate,
\begin{equation}
r_{lim} = r_0 \exp\big(\frac{1 - E^2}{\gamma\mu\varepsilon^2(2 - E^2)}\big)
\end{equation}
we have $v = 0$. However, the spatial region 
$r_{lim} \le r \le r_{0}$ can not be reached by the test particle, 
in its free motion. Hence, it can move only in the region 
$r_{0} \leq r\leq \infty $.

{\it c) Circular motion}. In this case $r = const$ and from eq. (15) 
we immediately have
$$\omega = r\frac{d\Phi}{dx^0} = \pm\big(\frac{E^2 - 1}{E^2}\big)^{1/2}\big(1 -$$
\begin{equation}
\frac{2 - E^2}{1 - E^2}\gamma\mu\varepsilon^2\ln\frac{r}{r_0}\big).
\end{equation}
This is the particle's circular velocity determined by clock of
infinitely far observer. 
For the local motionless observer, the circular velocity is, approximately, 
$$\Omega = r\frac{d\Phi}{d\tau} = r\frac{d\Phi}{dx^0}
(1 + \gamma\mu\varepsilon^2\ln\frac{r}{r_0}) = $$
\begin{equation}
\big(\frac{E^2 - 1}{E^2}\big)^{1/2}\big(1 - \frac{1}{1 - 
E^2}\gamma\mu\varepsilon^2\ln\frac{r}{r_0}\big).
\end{equation}
It is easy to see from the last two expressions that both of them 
tend to the newtonian value, 
$\Omega = \displaystyle\bigl(\frac{E^2 - 1}{E^2}\bigr)^{1/2}$, 
as $r \to r_0$. 
This means that the particle's circular velocity is staying constant 
in the flat space. 
Moreover, the circular velocity is equal to zero if 
$r =r_{lim}$; see eq. (39). 
However, the test particle cannot move in the finite region 
$r_{lim} \leq r \leq r_0$, in this case too.

\section{Conclusion}

We have examined the isotropic and nonisotropic geodesic lines in the
gravitational field of nonlinear cosmic string. Nonlinear terms in the equation
of motion imply some new dynamical effects, in both 
the light rays propagation and the test particle motion. We have found
additional deflection angle for the null and timelike geodesic lines; 
see eqs. (29) and (36). As to the timelike geodesics,
we have found equivalence of the radial and circular velocities, 
in the newtonian approximation.

We conclude this Section with the following remark, 
Dynamical processes in the vicinity of cosmic strings
are not limited to the light rays and test particle propagation. 
Research of the reciprocal dynamics of cosmic strings 
(oscillating cosmic strings [12]) 
is important in studying the early Universe cosmology.
So, the next step in investigation of the nonlinear cosmic 
strings interactions could be dynamics of the test thread, 
in the nonlinear cosmic string background.

\section*{Acknowledgment}

This work was done under the partial support within the program
"Organization and evolution of natural structures",
National Academy of Sciences, Kazakstan.


\newpage
\begin{thebibliography}{99}
\bibitem{L} A. D. Linde. {\it Particle Physics and Inflationary Cosmology}
(Harwood, Switzerland, 1990).
\bibitem{N} V. V. Novojilov. {\it Foundations of the Nonlinear Elastic Theory}
(Moscow, Gostechizdat, 1948) (in Russian);
            G. B. Whitham. {\it Linear and Nonlinear Waves}.
(New York, London, Sydney, Toronto, 1974).
\bibitem{C}  L. M. Chechin, T. B. Omarov. Hadronic J. {\bf 22} (1999) 197.
\bibitem{V}  A. Vilenkin. Phys. Rev. {\bf D 23} (1981) 852.
\bibitem{G}  J. R. Gott. Astrophys. J. {\bf 288} (1985) 422.
\bibitem{Ga} D. Garfinkle, T.Vachaspati. Phys. Rev. {\bf D 37} (1988) 2537.
\bibitem{Vo} D. N. Vollik, W.G.Unruh. Phys. Rev. {\bf D 42} (1990) 2621.
\bibitem{B}  A. Banerjee, N.Banerjee. Phys. Lett. {\bf A 160} (1991) 119.
\bibitem{Ch} S. Chakraborty, L.Biswas. Class. Quant. Grav. {\bf 13} (1996) 2153.
\bibitem{P}  P. Peter, D. Puy. Phys. Rev. {\bf D 48} (1993) 5546.
\bibitem{F}  V. Fock. {\it The Theory of Space, Time and Gravitation}
(Macmillan, New York, 1964).
\bibitem{O} T. B. Omarov, L. M. Chechin. Gen. Relat. Grav. {\bf 31} (1999) 443.
\end{thebibliography}
\end{document}